\documentclass[12pt,a4paper]{article}

\newcommand{\al}{\alpha'}
\newcommand{\e}{\epsilon}
\newcommand{\F}{\mathcal{F}}
\newcommand{\te}{\theta}
\newcommand{\Te}{\Theta}
\newcommand{\la}{\lambda}

\begin{document}

\begin{flushright}
hep-th/0003204
\end{flushright}
\vspace{1.8cm}

\begin{center}
\textbf{\Large Open String and Morita Equivalence \\
of the Dirac-Born-Infeld Action with Modulus $\Phi$}
\end{center}
\vspace{1.6cm}
\begin{center}
 Shijong Ryang
\end{center}

\begin{center}
\textit{Department of Physics \\ Kyoto Prefectural University of Medicine
\\ Taishogun, Kyoto 603-8334 Japan}
\par
\texttt{ryang@koto.kpu-m.ac.jp}
\end{center}
\vspace{2.8cm}
\begin{abstract}
Based on the canonical quantization of open strings ending on D-branes
with a background $B$ field, we construct the open string propagator.
We demonstrate the relation between the T duality of the
underlying string theory and the Morita equivalence of the 
interpolating general Dirac-Born-Infeld theory on a noncommutative torus
in the nonzero modulus $\Phi$ sector. The general noncommutative 
Dirac-Born-Infeld action with the Wess-Zumino terms expressed by the 
background R-R fields is shown to be Morita invariant.
\end{abstract}
\vspace{3cm}
\begin{flushleft}
March, 2000
\end{flushleft}
\newpage
Many insights of the noncommutativities in the string-M theories have 
been accumulated since the Yang-Mills theory on a noncommutative torus was
found to describe the DLCQ of M theory on a torus with a three-form field
background  \cite{CDS}. The Matrix theory description of the gauge 
theory on the noncommutative torus has been extensively studied 
\cite{NCY} to give the BPS energy spectrum and show the Morita 
equivalence of it \cite{PH,BM,KS} or of the action itself \cite{AS,BMZ},
which is argued to be related with the T-duality $SO(d,d,Z)$ of type II 
string theory compactified on the $d$-torus \cite{BMZ,PS}. The Morita 
equivalence of the BPS mass spectrum and its relation with the T-duality
have been also studied by a canonical description of the Dirac-Born-Infeld
(DBI) theory on a noncommutative torus \cite{HV}. 

On the other hand in the presence of a constant background NS-NS $B$ field
the world-volume of D-branes becomes noncommutative \cite{DH,AAS,CH}
and then the low energy dynamics of D-branes is naturally described by 
the noncommutative Yang-Mills (NCYM) theory. The quantization of open
strings propagating in the background $B$ field is directly related to the
noncommutativity of D-branes, where the world-volume coordinates of 
D-branes, which are in the static gauge identified with
 the space-time coordinates of open string 
end-points, are shown not to commute by means of the string mode expansion
method \cite{AAS,CH,FZ} and the Green function method \cite{SW}. 
The former method has been rigorously studied 
by Dirac's constrained quantization
procedure for the mixed boundary condition \cite{ACKL}.

Though there is in general a coupling between the open and closed strings,
we can take an appropriate low energy limit in such a way that the closed
strings decouple from the open strings ending on the D-branes
and the resulting theory for the open strings reduces to
the NCYM theory \cite{SW}. Specially through the different 
regularizations the equivalence between the ordinary Yang-Mills theory 
and the NCYM theory has been shown by comparing the ordinary DBI theory
with the noncommutative DBI theory where the background $B$ field is 
replaced by the noncommutativity of the world-volume coordinates.
Moreover a general DBI theory interpolating the two theories has been 
proposed to be described by a noncommutative action including an extra 
modulus $\Phi$, which is known to appear as a magnetic background
in the NCYM action and show a
particular transformation under the group of the Morita equivalence for
the NCYM theory \cite{KS,PS}. Based on this general DBI theory the 
Morita transformation rules have been reproduced through the appropriate
low energy limit from the T-duality of type II string theory in the zero
$\Phi$ sector.

We will try to extend the results of Seiberg and Witten \cite{SW} for 
the interrelations among the string dynamics, the noncommutative DBI 
theory and the NCYM theory. In order to see whether the canonical
quantization of open strings ending on D-branes in the presence of a 
constant background $B$ field \cite{CH} 
is consistently well defined or not, we will calculate the string 
propagator and compare with that derived from
the Green function method. Based on the interpolating DBI theory on a 
noncommutative torus we will study how the Morita equivalence 
transformation is related with the T-duality generally in the nonzero
$\Phi$ sector. The invariance of 
the general noncommutative DBI action with
modulus $\Phi$ and the Wess-Zumino (WZ) action themselves under 
the Morita equivalence transformation will be investigated and 
the transformation properties of the  background R-R fields will 
be elucidated.

The bosonic part of the classical action for an open string ending on a 
D$p$-brane is given by
\begin{equation}
S = \frac{1}{4\pi \al}\int_{\Sigma}d^2\sigma (g_{\mu\nu}\partial_aX^{\mu}
\partial^aX^{\nu} - 2\pi \al B_{\mu\nu}\e^{ab}\partial_aX^{\mu}\partial_b
X^{\nu})+\oint_{\partial\Sigma}d\tau A_i(X)\partial_{\tau}X^i,
\end{equation}
where $A_i, i = 0,1,\cdots,p$ is the U(1) gauge field living on the 
D$p$-brane and $g_{\mu\nu}, B_{\mu\nu}, \mu = 0,1,\cdots,9$ represent the
constant background metric and the constant background NS-NS two-form 
field.  Since the background $B_{\mu\nu}$ field can have nonzero 
components only along the directions parallel to the D$p$-brane, the 
action can be expressed as 
\begin{equation}
S = \frac{1}{4\pi \al}\int_{\Sigma}d^2\sigma (g_{\mu\nu}\partial_aX^{\mu}
\partial^aX^{\nu} - 2\pi \al \F_{ij}\e^{ab}\partial_aX^{i}\partial_b
X^{j})
\end{equation}
in terms of the gauge invariant field strength 
$\F_{ij} = B_{ij} + F_{ij}$ with $F = dA$. The transverse string
variable $X^{\mu}, \mu = p+1,\cdots,9$ can be treated trivially so
that we will be concerned with the longitudinal variable $X^i$ only.
The solution of $X^i$ to the equation of motion, satisfying the mixed
boundary condition $\partial_{\sigma}X^i - 2\pi\al\partial_{\tau}
X^j\F_j^i = 0$ at $\sigma = 0, \pi$, is given by  
\begin{equation}
X^k = x_0^k + ( p_0^k\tau + 2\pi\al p_0^j\F_j^k\sigma )  
 + \sum_{n\neq 0}\frac{e^{-in\tau}}{n}(ia_n^k \cos n\sigma + 2\pi\al
a_n^j\F_j^k \sin n\sigma).
\label{mod}\end{equation}
In Ref. \cite{CH} the quantization of $X^k$ was performed by a 
generalization of the canonical approach analyzing the 
time-averaged symplectic form on the phase space. The commutation 
relations for the modes were extracted as 
\[ [a_m^i, a_n^j] = 2\al mM^{-1ij}\delta_{m+n}, \hspace{1cm} 
[x_0^i, p_0^j] =i2\al M^{-1ij}, \]
\begin{equation}
 [x_0^i, x_0^j] = i2\pi \al(M^{-1}\F)^{ij},  \hspace{1cm}
 [p_0^i, p_0^j] = 0,
\label{com}\end{equation}
where $M_{ij} = g_{ij} - (2\pi\al)^2(\F g^{-1}\F)_{ij}$ and 
$(M^{-1}\F)^{ij}$ is abbreviation for $M^{-1ik}\F_{kl}g^{-1lj}$.

Making use of these results we construct the propagator of 
the open string position operator $X^k(\sigma,\tau)$. 
 Here we turn to the Euclidean metric on the world 
sheet. By substituting $\tau = -i\tau'$ into (\ref{mod}) and using $z =
e^{\tau'+i\sigma}$ we have the following normal mode expression
\begin{equation}
X^k(z) = x_0^k - \frac{i}{2}(p_0^k\ln z\bar{z} + 2\pi\al p_0^j\F_j^k
\ln \frac{z}{\bar{z}}) + i\sum_{n\neq 0} (a_n^k(z^{-n} + \bar{z}^{-n}) 
+ 2\pi\al a_n^j\F_j^k(z^{-n} - \bar{z}^{-n})).
\end{equation}
The commutation relations in (\ref{com}) provide the string propagator
\begin{eqnarray}
<0|X^i(z)X^j(z')|0> = \al ( -M^{-1ij}\ln z\bar{z} +2\pi\al(\F M^{-1})^{ij}
\ln \frac{z}{\bar{z}} \hspace{3cm} {}\nonumber \\
-\frac{1}{2}M^{-1ij}(\ln |1-\frac{z'}{z}|^2 + \ln 
|1-\frac{\bar{z}'}{z}|^2) + \frac{(2\pi\al)^2}{2}(\F M^{-1}\F)^{ij}(\ln 
|1-\frac{z'}{z}|^2 - \ln |1-\frac{\bar{z}'}{z}|^2) \\
- \frac{2\pi\al}{2}(M^{-1}\F)^{ij}\ln \frac{(z-z')(\bar{z}-z')}
{(z-\bar{z}')(\bar{z}-\bar{z}')} + \frac{2\pi\al}{2}(\F M^{-1})^{ij}(\ln
\frac{(z-z')(z-\bar{z}')}{(\bar{z}-z')(\bar{z}-\bar{z}')} - 
2\ln \frac{z}{\bar{z}})), \nonumber
\end{eqnarray}
whose first and second terms are the zero mode contributions. Owing to
 the relations $M^{-1} - (2\pi\al)^2g^{-1}\F M^{-1}\F g^{-1} = g^{-1}, 
M^{-1}\F g^{-1} = g^{-1}\F M^{-1}$, where abbreviation is not used but the
indeces are raised by metric, the propagator turns out to be
\begin{equation}
-\al(g^{-1ij}(\ln|z-z'|-\ln|z-\bar{z}'|) + M^{-1ij}\ln|z-z'|^2
-2\pi\al(M^{-1}\F g^{-1})^{ij}\ln\frac{z-\bar{z}'}{\bar{z}-z'}).
\label{pro}\end{equation}
In view of $M^{-1} = (g+2\pi\al \F)^{-1}g(g-2\pi\al\F)^{-1}$ and 
$M^{-1}\F g^{-1} = (g+2\pi\al\F)^{-1}\F(g-2\pi\al\F)^{-1}$ we note that we
have obtained the same expression as that mentioned in Ref. \cite{SW},
where the open string propagator is constructed as a solution to 
the equation of motion for the Green function satisfying the 
mixed boundary condition.

Seiberg and Witten have proposed a general DBI theory with an arbitrary
$\te$ and parameters $G, \Phi$ determined by a formula
\begin{equation}
\frac{1}{G + 2\pi\al \Phi} = - \frac{\te}{2\pi\al} + 
\frac{1}{g + 2\pi\al B},
\label{swf}\end{equation}
which implies that the open string metric $G$ and the antisymmetric 
two-form $\Phi$ are expressed in terms of the closed string parameters
$g, B$ and the noncommutative parameter $\te$ \cite{SW}. 
In the formula (\ref{swf}) expressed by the $(p+1)\times (p+1)$ matrices
we assume that $B_{0i}=0$ as well as $g_{0i}=0$, here $i=1,\cdots,p$ 
in the Lorentzian target space-time. 
For the D$p$-brane compactified on a $p$-torus $T^p$
parametrized by $x^i \sim x^i + 2\pi r$,
$i = 1,\cdots,p$ with closed string metric $g_{ij}$, the T-duality
$SO(p,p,Z)$ transformation on $E = r^2(g + 2\pi\al B)/\al$ is given by 
$E' = (aE + b)(cE + d)^{-1}$, with $c^ta + a^tc = 0, d^tb + b^td = 0,
c^tb + a^td = 1$ where $a, b, c$ and $d$ are $p\times p$ matrices with
integer entries. We consider the case that the $\te$ dependence appears
through the noncommutativity of the world-volume space-coordinates, 
so that  $\te^{0i}$ is set to zero.
 Therefore we have a formula (\ref{swf}) expressed by the
$p\times p$ matrices, $G_{00}=g_{00}$ and 
$\Phi_{0i}=0$ which corresponds to the treatment 
of $\Phi$ as the magnetic background. From the $p\times p$ 
matrix formula (\ref{swf})  using the dimensionless $p\times p$ 
matrix $\Te = \te/2\pi r^2$ we extract $G$ and $\Phi$ as
\begin{eqnarray}
G &=& \frac{\al}{2r^2} \frac{1}{1+E^t\Te}( E + E^t )\frac{1}{1-\Te E}, 
\nonumber \\ 
\Phi &=& \frac{1}{4\pi r^2} \frac{1}{1+E^t\Te}( 2E^t\Te E + E - E^t )
\frac{1}{1-\Te E}.
\label{gph}\end{eqnarray}
Compared with the given closed string parameters $g$ and $B$, the NCYM
theory includes the corresponding two parameters $G$ and $\te$ with an 
extra modulus $\Phi$ added, which is associated with $SO(p)$ symmetry for
the relativistic generalization of some NCYM energy that is identified 
with the BPS mass formula of the noncommutative DBI theory \cite{PS}.
Here assuming that $\Te$ transforms by a fractional transformation as
$\Te \rightarrow \Te' = (c + d\Te)(a + b\Te)^{-1}$ we  examine
how $G$ and $\Phi$ transform with respect to the T-duality. There are 
interesting symmetric relations, $E' + E'^t = (E^tc^t + d^t)^{-1}
(E + E^t)(cE+ d)^{-1}$ and 
\begin{equation}
1- \Te 'E' = \frac{1}{a^t - \Te b^t}(1-\Te E)\frac{1}{cE + d}.
\label{the}\end{equation}
Combining them with the first equation of (\ref{gph}) and ${\Te'}^t =
-\Te'$ we derive the transformation for the open string metric
\begin{equation}
G' = ( a + b\Te ) G ( a + b \Te )^t.
\label{tg}\end{equation}

Similarly the transformed modulus $\Phi'$ is represented by 
\begin{equation}
\Phi' = \frac{1}{4\pi r^2}(a+b\Te)\frac{1}{1+E^t\Te}(2X+Y)
\frac{1}{1-\Te E}(a+b\Te)^t,
\end{equation}
where $X = (E^ta^t+b^t)(c+d\Te)(a+b\Te)^{-1}(aE+b)$ and $Y$ is given by
$Y = E-E^t + 2Y_0$ with $Y_0 = -(E^ta^t + b^t)cE + (E^tc^t+d^t)b$.
For convenience, $X=(E^ta^t + b^t)\Te'(aE + b)$ is equivalently 
rewritten by
\begin{equation}
X = \frac{1}{2}(E^ta^t+ b^t)((c+d\Te)\frac{1}{a+b\Te}- 
\frac{1}{a^t-\Te b^t}(c^t-\Te d^t))(aE + b)
\end{equation}
because of  ${\Te'}^t = - \Te'$, which is further expressed as
$X=( E^t + (1+E^t\Te)b^t(a^t-\Te b^t)^{-1})(\Te +X_0)(E+ (a+b\Te)^{-1}
b(1-\Te E))$ with $X_0 = -(c^t-\Te d^t)b\Te + (a^t - \Te b^t)c$.
From these expressions we 
first extract some terms suggested by the second equation in (\ref{gph})
as $2X + Y = 2E^t\Te E + E - E^t + 2Z,$
where the remaining terms are gathered by
\begin{eqnarray}
Z &=& E^tX_0E + E^t(\Te + X_0)\frac{1}{a + b\Te} b(1-\Te E) \nonumber \\
 &+& (1 + E^t\Te)b^t\frac{1}{a^t -\Te b^t}(\Te + X_0)E + Y_0 + Z_0
\label{zyz}\end{eqnarray}
with $Z_0 = (1+ E^t\Te)b^t(a^t- \Te b^t)^{-1}
(\Te+X_0)(a+b\Te)^{-1}b(1-\Te E)$.

The substitution of an identity
\begin{equation}
c^t - \Te d^t = - (a^t - \Te b^t)(c + d\Te)\frac{1}{a + b\Te}
\label{cd}\end{equation}
into $X_0$ in (\ref{zyz}) brings two kinds of cancellations in 
the first term against the second and third terms. As a result 
the first term of (\ref{zyz}) becomes
$E^t(a^tc+(a^t-\Te b^t)d\Te(a+b\Te)^{-1}b\Te)E$, whose first term is
further cancelled against a term in $Y_0$. Since there is 
another cancellation between the second and third terms
through $b^t(c+d\Te)(a+b\Te)^{-1} = (a+b\Te)^{-1} - d^t$,
the resulting third term can be expressed as $[(1+E^t\Te)
(b^t(a^t-\Te b^t)^{-1} - d^tb) + (a+b\Te)^{-1}b]\Te E + b^tcE$, 
whose last term is also cancelled against a term in $Y_0$.
The remainging second term is given by $E^t[(\Te + (a^t-\Te b^t)c)
(a+b\Te)^{-1}b - (c^t-\Te d^t)b\Te(a+b\Te)^{-1}b(1-\Te E)]$,
which is further changed into $E^t[(a^tc + \Te d^ta)(a+b\Te)^{-1}b
-(c^t-\Te d^t)(1-a(a+b\Te)^{-1})b(1-\Te E)]$ where a combination 
$E^t(a^tc+c^ta)(a+b\Te)^{-1}b$ vanishes. Thus we have
\begin{eqnarray}
Z &=& E^t(c^tb + (a^t-\Te b^t)(c + d\Te)\frac{1}{a+b\Te}b + 
\Te\frac{1}{a+b\Te}b -2\Te d^tb)\Te E \nonumber \\
&+& E^t\Te d^tb + (\frac{1}{a+b\Te} - d^tb)\Te E + d^tb + Z_0
+ (1+ E^t\Te)b^t\frac{1}{a^t -\Te b^t}\Te E,
\label{zz}\end{eqnarray}
whose last term is cancelled against a term in $Z_0=(E^t\Te +1)b^t
(a^t-\Te b^t)^{-1}(\Te +X_0)(-E + (a+b\Te)^{-1}(aE+b))$.
The second term in (\ref{zz}) is simplified into $-E^t(c^t-\Te d^t)b
\Te E$ through (\ref{cd}). Then the remaining $Z_0$ is further arranged
by using (\ref{cd}) for $X_0$ into
\begin{eqnarray}
Z_0 &=& (E^t\Te +1)b^t[ \frac{1}{a^t-\Te b^t}\Te\frac{1}{a+b\Te}(aE+b)
\nonumber \\ &+& ((c+d\Te)\frac{1}{a+b\Te}b\Te + c)\frac{1}{a+b\Te}
b(1-\Te E)].
\label{zo}\end{eqnarray}
Collecting the three terms with $b$ on the right end in (\ref{zo}) and
making use of 
\begin{equation}
\frac{1}{a^t - \Te b^t}\Te + c = (c + d\Te)\frac{1}{a + b\Te}a,
\label{ab}\end{equation}
we can see that they are simply expressed as $(E^t\Te +1)b^t(c +d\Te)
(a+b\Te)^{-1}b$. This compact expression further takes the form
$(E^t\Te + 1)(a +b\Te)^{-1}b -E^t\Te d^tb -d^tb$ whose last two terms are
cancelled out in $Z$ (\ref{zz}). The three terms with $E$ on the right 
end in (\ref{zo}) turn out to be $(E^t\Te +1)(d^tb - 2(a+b\Te)^{-1}b)
\Te E$ also through (\ref{ab}). Gathering together we arrive at a
simplified expression
\begin{equation}
Z = (E^t\Te + 1)\frac{1}{a + b\Te} b(1 - \Te E),
\end{equation}
which yields
\begin{equation}
\Phi' = (a + b\Te)\Phi(a + b\Te)^t + \frac{1}{2\pi r^2}b(a + b\Te)^t.
\label{phi}\end{equation}

The $SO(p,p,Z)$ T-duality action on the closed string coupling $g_s$ is
given by $g'_s = g_s/\det(cE+d)^{1/2}$. The effective open string 
coupling $G_s = g_s(\det(G+2\pi \al \Phi)/\det(g+2\pi\al B))^{1/2}$ 
specified by the $(p+1)\times(p+1)$ matrices is expressed as
$G_s  = g_s(\det(1-\Te E))^{-1/2}$ since $G_{00}=g_{00}, \Phi_{0i}
=B_{0i}=0$.  Therefore the relation (\ref{the}) yields the
$SO(p,p,Z)$ T-duality action on $G_s$
\begin{equation}
G_s' = \sqrt{\det(a + b\Te)}G_s,
\label{gs}\end{equation}
which further determines the transformation for the Yang-Mills gauge
coupling $g_{YM} =((2\pi)^{p-2}G_s/(\al)^{(3-p)/2})^{1/2}$ to be 
\begin{equation}
g'_{YM} = g_{YM}(\det(a + b\Te))^{\frac{1}{4}}.
\label{gym}\end{equation}
The obtained expressions such as (\ref{tg}), (\ref{phi}) and (\ref{gym})
are just the Morita transformation rules in the NCYM theory with modulus
$\Phi$. In the $\Phi = 0$ orbit the T-duality action on the closed string
metric $g$ and NS-NS $B$ field was shown to be mapped to the Morita 
transformation only when the zero slope limit was taken in such a way that
$E^{-1} \approx \Te$ \cite{SW}. In our general nonzero $\Phi$ case this 
mapping naturally appears where we do not restrict ourselves to the zero
slope limit. We would like to interpret this general mapping as a direct
relation between  the Morita equivalence of the 
noncommutative DBI theory and the T-duality.

In order to confirm this interpretation we apply the above Morita 
transformation rules  to the non-abelian DBI theory
living on a noncommutative torus. The effective action of D$p$-branes
with the modulus two-form $\Phi$ on a $p$-dimensional noncommutative torus
is given by 
\begin{equation}
S = - \frac{1}{G_s(2\pi)^p\al^{\frac{p+1}{2}}}\int d^{p+1}\sigma 
Tr_{\te}\sqrt{-\det(G + 2\pi\al(F + \Phi))} + S_{WZ}.
\end{equation}
The first term is regarded as an general DBI action interpolationg between
the ordinary DBI one on a commutative torus with $G=g, \Phi=B, G_s=g_s$
and the commutative gauge fields, and the noncommutative DBI one 
expressed in terms of the open string variables with $\Phi=0$ and
the noncommutative gauge fields \cite{SW}. The $Tr_{\te}$
is the trace on the gauge bundle on the noncommutative torus and
is regarded as the symmetric trace \cite{AT} for the non-abelian group. 
We assume that the WZ action 
$S_{WZ}$ on a noncommutative torus takes the same form as that on a 
commutative torus \cite{LDS}. So it is represented by 
$S_{WZ}=\int Tr_{\te}(\sum C^{(n)})e^{2\pi\al F}$ in terms of a pullback
of the $n$-form R-R potential $C^{(n)}$, where on the exponential $F$ 
only appears and the background $B$ field is replaced by the 
noncommutativity $\te$. In the DBI action the square root part is 
decomposed into $\sqrt{\det(G+2\pi\al(F+\Phi))_{ij}} (-(G_{00}-
(2\pi\al)^2 F_{0i}(G+ 2\pi\al(F+\Phi))^{-1ij}F_{j0}))^{1/2}$, 
where $i=1,\cdots,p$ and we have taken account of
$\Phi_{0i}=G_{0i}=0$. Under the Morita equivalence
transformation on a $p$-dimensional noncommutative torus the magnetic
component of the gauge field strength $F_{ij}$ is transformed as
$F_{ij} \rightarrow ((a+b\Te)F(a+b\Te)^t -\frac{1}{2\pi r^2}b
(a+b\Te)^t)_{ij}$ according to (\ref{phi}), while the electric component
$F_{0i}$ is transformed as
$F_{0i} \rightarrow (F(a+b\Te)^t)_{0i}$.  
These transformation rules are argued in the investigation of the Morita
equivalence in the context of the NCYM theory \cite{BM,KS,AS,BMZ}.
We can take $G_{00}=g_{00}$ to be unchanged since the T-duality 
transformation is performed on the directions of the $p$-torus.
The transformation of the trace $Tr_{\te}$
is given by $Tr_{\te} \rightarrow (\det(a+b\Te))^{-1/2}Tr_{\te}$
\cite{HV}. Under these transformation rules together with (\ref{tg}),
(\ref{phi}) and(\ref{gs}) we can see that the general interpolating DBI
action is invariant.

For the WZ action we consider the $D5$-brane theory for concreteness and
write down
\begin{eqnarray}
S_{WZ} &=& \int d^6\sigma Tr_{\te}\e^{ilklm}(\frac{2\pi\al}{24}F_{0i}
C_{jklm} + \frac{(2\pi\al)^2}{4}F_{0i}F_{jk}C_{lm} + 
\frac{(2\pi\al)^3}{8}F_{0i}F_{jk}F_{lm}C \nonumber \\  
&+& \frac{1}{5!}C_{0ijklm}+ \frac{2\pi\al}{12}C_{0ijk}F_{lm} +  
\frac{(2\pi\al)^2}{8}C_{0i}F_{jk}F_{lm}).
\label{swz}\end{eqnarray}
The WZ action for the first three terms is shown to be invariant
under the Morita equivalence transformation, if we take the R-R 
potentials such as $C, C_{lm}$ and $C_{jklm}$ to transform 
simultaneously as
\begin{eqnarray}
C &\rightarrow& \frac{1}{\sqrt{\det A}}C, \hspace{1cm}
 C_{ij} \; \rightarrow  \; \frac{1}
{\sqrt{\det A}}((ACA^t)_{ij} + \frac{\al}{r^2}(bA^t)_{[ij]}C), 
\\  C_{ijkl} &\rightarrow& 
\frac{1}{\sqrt{\det A}}(A_{[i}^a A_j^b A_k^c A_{l]}^d
C_{abcd} + \frac{6\al}{r^2}A_{[i}^a A_j^b (bA^t)_{kl]}C_{ab} +
3(\frac{\al}{r^2})^2 C(bA^t)_{[ij}(bA^t)_{kl]} ), \nonumber
\end{eqnarray}
where $A = a + b\Te$ and $A_i^a$ stands for $A_i^{\:\:a}$ and the
 appropriate antisymmetrization factors 6, 3 appear.
The invariance of $S_{WZ}$ for the last three terms 
is also shown, if we make $C_{0i}, C_{0ijk}$ and $C_{0ijklm}$ transform
in the following way
\begin{eqnarray}
C_{0i} &\rightarrow&  \frac{1}{\sqrt{\det A}}A_i^a C_{0a}, \hspace{1cm}
C_{0ijk} \;\rightarrow \;\frac{1}{\sqrt{\det A}}(A_{[i}^aA_j^b A_{k]}^c
C_{0abc} + \frac{3\al}{r^2} A_{[i}^a(bA^t)_{jk]}C_{0a} ), \nonumber \\
C_{0ijklm} &\rightarrow& \frac{1}{\sqrt{\det A}}(A_{[i}^a \cdots A_{m]}^e
C_{0a\cdots e} + \frac{10\al}{r^2}A_{[i}^a A_j^b A_k^c (bA^t)_{lm]}
C_{0abc} \\  &+& 15(\frac{\al}{r^2})^2
A_{[i}^a(bA^t)_{jk}(bA^t)_{lm]}C_{0a} ), \nonumber
\end{eqnarray}
where the coefficients 3, 10 and 15 also show the appropriate 
antisymmetrization factors specified by ${}_3C_1, {}_5C_3$ and 
$3\cdot{}_5C_1$.

Redefinitions of the above R-R  potentials as 
\begin{eqnarray}
\la^i = \frac{1}{4!}\e^{ijklm}C_{jklm}, \; \la^{ijk} = \frac{1}{2!}
\e^{ijklm}C_{lm}, \; \la^{ijklm}=\e^{ijklm}C, \nonumber \\
\la =\frac{1}{5!}\e^{ijklm}C_{0ijklm},\; \la^{ij} = \frac{1}{3!}
\e^{ijklm}C_{0klm}, \; \la^{ijkl}=\e^{ijklm}C_{0m}
\label{ram}\end{eqnarray}
enable us to write the WZ action (\ref{swz}) to be
\begin{eqnarray}
S_{WZ} &=& \int d^6\sigma Tr_{\te}(2\pi\al F_{0i}\la^i + 
\frac{(2\pi\al)^2}{2}F_{0i}F_{jk}\la^{ijk} + \frac{(2\pi\al)^3}{8}
F_{0i}F_{jk}F_{lm}\la^{ijklm} \nonumber \\
&+& \la + \frac{2\pi\al}{2}F_{ij}\la^{ij} + \frac{(2\pi\al)^2}{8}
F_{ij}F_{kl}\la^{ijkl} ).
\end{eqnarray}
The first three terms are also invariant under the Morita transformation
accompanied with 
\begin{eqnarray}
\la^i &\rightarrow& \sqrt{\det A}(B_a^{i}\la^a + \frac{\al}{2r^2}
B_a^iB_b^jB_c^k\la^{abc}(bA^t)_{jk}) + \frac{1}{8} (\frac{\al}{r^2})^2
\frac{1}{\sqrt{\det A}}\la^{ijklm}(bA^t)_{jk}(bA^t)_{lm}, \nonumber \\
\la^{ijk} &\rightarrow& \sqrt{\det A}B_a^i B_b^j B_c^k \la^{abc} +
\frac{\al}{2r^2} \frac{1}{\sqrt{\det A}}\la^{ijklm}(bA^t)_{lm}, \;
\la^{ijklm} \rightarrow \frac{1}{\sqrt{\det A}}\la^{ijklm},
\label{tra}\end{eqnarray}
which are obtained from (25) by defining $B =(A^{-1})^t$ and
using $B_a^i$ for $B_{\:\:a}^i$. The
invariance for the last three terms is also provided by the
following transformation properties of the redefined R-R potentials
\begin{eqnarray}
\la &\rightarrow& \sqrt{\det A}(\la + \frac{\al}{2r^2}B_a^i B_b^j
(bA^t)_{ij}\la^{ab}  + \frac{1}{8}
(\frac{\al}{r^2})^2B_a^i B_b^j B_c^k B_d^l
(bA^t)_{ij}(bA^t)_{kl}\la^{abcd}),  \\
\la^{ij} &\rightarrow& \sqrt{\det A}(B_a^i B_b^j \la^{ab} + 
\frac{\al}{2r^2}B_a^i B_b^j B_c^k B_d^l(bA^t)_{kl}\la^{abcd}),\;
\la^{ijkl} \rightarrow \sqrt{\det A}B_a^i B_b^j B_c^k B_d^l\la^{abcd},
\nonumber
\end{eqnarray}
which are derived by combining (26) and (\ref{ram}). 
Thus we have observed the Morita equivalence of the action itself
for the interpolating general DBI theory, which is compared with
the Morita equivalence of the BPS energy spectrum  based on the canonical
approaches of the DBI theory on the noncommutative two- or four- torus
\cite{HV} and the NCYM theory \cite{PH,BM,KS}.
The action of the Yang-Mills theory on a noncommutative three-torus with
the Chern-Simon topological terms is proved to be Morita invariant in
Ref.\cite{BMZ}. The transformation behaviors of the zero-form and two-form
R-R potentials given there are now extended to (25) for the 
noncommutative five-torus, where the transformation property of the
four-form R-R potential is obtained in a suggestive and systematic
form, while the corresponding transformation rules (\ref{tra}) for the 
redefined R-R potentials are considered as an extension of those for 
the noncommutative three-torus presented in Ref. \cite{KS}.
In Refs. \cite{KS,BMZ} only the space components of the R-R potentials
are assumed to be nonzero. Here even if we take account of the R-R 
potentials with a time component the Morita equivalence of the WZ action
for them is confirmed to hold separately, where the transformation of 
the six-form R-R potential is specified.

We have shown that the quantization of open strings ending on the 
D-branes with the background $B$ field is so well constructed
from the symplectic structure on the phase space that
we can consistently derive the string propagator with the mixed boundary
conditions. Combining the T-duality transformation for the closed string
parameters with the fractional transformation for the noncommutativity
parameter we can extract the Morita transformation rules for the open
string parameters which characterize the interpolating noncommutative
DBI action, without recourse to the low energy zero slope limit.
The interpolating DBI action with nonzero modulus $\Phi$ as well as
the WZ action on a noncommutative torus have been demonstrated to have
the same Morita equivalence as the NCYM theory. Therefore 
it can be said that the T-duality transformation is 
directly translated to the Morita equivalence in the context of 
the noncommutative DBI theory, which is considered as the generalization
of the indirect low-energy extraction of the Morita equivalence
for the NCYM theory. 

On the five-dimensional noncommutative torus
we have seen that the transformed R-R potentials are expanded in 
terms of the same and lower rank R-R potentials with the appropriate
coefficients which appear as the antisymmetrization numbers.
This expansion reflects the characteristics for the couplings of the
D$p$-brane not only to the the R-R potential with the $(p+1)$ rank 
but also to the R-R potentials with the ranks lower 
by even numbers, accompanied with the additional and
multiple interactions to the world-volume gauge field strength
in the noncommutative WZ acion. 
We speculate that this expansion is related with the recent
works about the noncommutative description of D-branes 
\cite{ICSK} where the D-branes can be expressed as a configuration of
infinitely many lower dimensional D-branes.
It is tempting to suspect
that the application of the Morita transformation of the
noncommutative DBI theory is useful to build the nontrivial
bound state consisting of a D$p$-brane and the D-branes with lower 
world-volume dimensions from a pure D$p$-brane configuration.


\begin{thebibliography}{99}
\bibitem{CDS} A. Connes, M.R. Douglas and A. Schwarz, JHEP 
\textbf{02} (1998) 003, hep-th/9711162.
\bibitem{NCY} P.-M. Ho, Y.-Y. Wu and Y.-S. Wu, Phys. Rev. \textbf{D58}
(1998) 026006, hep-th/9712201; M. Li, hep-th/9802052;
T. Kawano and K. Okuyama, Phys. Lett. \textbf{B433} (1998) 29,
hep-th/9803044; B. Morariu and B. Zumino, hep-th/9807198.
\bibitem{PH} P.-M. Ho, Phys. Lett. \textbf{B434} (1998) 41, hep-th/9803166.
\bibitem{BM} D. Brace and B. Morariu, JHEP \textbf{02} (1999) 004,
 hep-th/9810185.
\bibitem{KS} A. Konechny and A. Schwarz, Nucl. Phys. \textbf{B550} (1999)
561, hep-th/9811159; Phys. Lett. \textbf{B453} (1999) 23, hep-th/9901077.
\bibitem{AS} A. Schwarz, Nucl. Phys. \textbf{B534} (1998) 720, 
hep-th/9805034.
\bibitem{BMZ} D. Brace, B. Morariu and B. Zumino, Nucl. Phys. 
\textbf{B545} (1999) 192, hep-th/9810099; Nucl. Phys. 
\textbf{B549} (1999) 181, hep-th/9811213.
\bibitem{PS} B. Pioline and A. Schwarz, JHEP \textbf{08} (1999)
021, hep-th/9908019.
\bibitem{HV} C. Hofman and E. Verlinde, JHEP \textbf{12} (1998) 010,
hep-th/9810116;  Nucl. Phys. \textbf{B547} (1999) 157, hep-th/9810219.
\bibitem{DH} M.R. Douglas and C. Hull, JHEP \textbf{02} (1998) 008, 
hep-th/9711165.
\bibitem{AAS} F. Ardalan, H. Arfaei and M.M. Sheikh-Jabbari, 
hep-th/9803067;  JHEP \textbf{02} (1999) 016, hep-th/9810072.
\bibitem{CH} C.-S. Chu and P.-M. Ho, Nucl. Phys. \textbf{B550} (1999) 
151, hep-th/9812219.
\bibitem{FZ} A. Fayyazuddin and M. Zabzine, hep-th/9911018.
\bibitem{SW} N. Seiberg and E. Witten, JHEP \textbf{09} (1999) 
032, hep-th/9908142.
\bibitem{ACKL} F. Ardalan, H. Arfaei and M.M. Sheikh-Jabbari, 
hep-th/9906161; C.-S. Chu and P.-M. Ho, hep-th/9906192;
W.T. Kim and J.J. Oh, hep-th/9911085; T. Lee, hep-th/9911140.
\bibitem{AT} A.A. Tseytlin, Nucl. Phys. \textbf{B501} (1997) 41,
hep-th/9701125.
\bibitem{LDS} M. Li, Nucl. Phys. \textbf{B460} (1996) 351, 
hep-th/9510161; M.R. Douglas, hep-th/9512077; 
C. Schmidhuber, Nucl. Phys. \textbf{B467} (1996) 146, hep-th/9601003.
\bibitem{ICSK} N. Ishibashi, Nucl. Phys. \textbf{B539} (1999) 107,
hep-th/9804163; hep-th/9909176; L. Cornalba and R. Schiappa,
hep-th/9907211; L. Cornalba, hep-th/9909081;
T. Kuroki, hep-th/0001011.
\end{thebibliography}
\end{document}